\documentclass[conference]{IEEEtran}
\def\BibTeX{{\rm B\kern-.05em{\sc i\kern-.025em b}\kern-.08em T\kern-.1667em\lower.7ex\hbox{E}\kern-.125emX}}
\usepackage{amsmath,amssymb,amsmath,amsfonts}
\usepackage{bm}
\usepackage{bbm}
\usepackage{graphicx}
\usepackage{cite}
\usepackage{epstopdf}
\usepackage{gensymb}
\usepackage{color}
\usepackage{lipsum}
\usepackage{float}
\usepackage{epstopdf}
\usepackage[mathcal]{eucal}
\usepackage{subfiles}
\usepackage[T1]{fontenc}
\usepackage{siunitx}
\usepackage{tabulary}

\usepackage{epsfig,graphics,graphicx,subfig,amssymb,amstext,amsmath,algorithm,algorithmic,wrapfig,multirow}
\usepackage{array}
\usepackage{adjustbox}
\usepackage[dvipsnames]{xcolor}
\usepackage{cite}
\usepackage{times}
\usepackage{siunitx}

\usepackage{lipsum}
\usepackage{multicol}
\usepackage{mwe}

\usepackage{placeins}
\usepackage{textcomp}
\usepackage[font=small]{caption}
\usepackage{flushend}
\usepackage{xcolor}
\usepackage{tabularx}
\usepackage{bm}
\usepackage{enumerate}
\usepackage{tikz}
\usepackage{pgfplots}
\usepackage{epstopdf}
\usetikzlibrary{shapes,arrows}
\usepackage[utf8]{inputenc}
\usepackage{mathtools}
\usepackage[utf8]{inputenc}
\usepackage{xcolor}
\usepackage{tikz}
\usetikzlibrary{chains,arrows,calc,positioning}

\pgfplotsset{compat=1.16}
\definecolor{bittersweet}{rgb}{1.0, 0.44, 0.37}
\definecolor{glaucous}{rgb}{0.38, 0.51, 0.71}
\definecolor{gainsboro}{rgb}{0.86, 0.86, 0.86}
\definecolor{babyblueeyes}{rgb}{0.63, 0.79, 0.95}
\definecolor{silver}{rgb}{0.75, 0.75, 0.75}
\definecolor{neoncarrot}{rgb}{1.0, 0.64, 0.26}

\usepackage{soul}

\usepackage{breqn}
\usepackage{booktabs}
\usepackage{multirow}
\usepackage{verbatim}
\usepackage{enumitem}
\usepackage{tabulary}

\usepackage[normalem]{ulem}

\usepackage[acronyms,nonumberlist,nopostdot,nomain,nogroupskip]{glossaries}
\newacronym{quic}{QUIC}{Quick UDP Internet Connections}
\newacronym{3gpp}{3GPP}{3rd Generation Partnership Project}
\newacronym{adc}{ADC}{Analog to Digital Converter}
\newacronym{5g}{5G}{5th generation}
\newacronym{aimd}{AIMD}{Additive Increase Multiplicative Decrease}
\newacronym{am}{AM}{Acknowledged Mode}
\newacronym{amc}{AMC}{Adaptive Modulation and Coding}
\newacronym{aqm}{AQM}{Active Queue Management}
\newacronym{awgn}{AGWN}{Additive White Gaussian Noise}
\newacronym{afd}{AFD}{Austin Fire Department}
\newacronym{balia}{BALIA}{Balanced Link Adaptation}
\newacronym{bdp}{BDP}{Bandwidth-Delay Product}
\newacronym{bf}{BF}{Beamforming}
\newacronym{cc}{CC}{Congestion Control}
\newacronym{cdf}{CDF}{Cumulative Distribution Function}
\newacronym{cn}{CN}{Core Network}
\newacronym{cqi}{CQI}{Channel Quality Information}
\newacronym{cp}{CP}{Control Plane}
\newacronym{csirs}{CSI-RS}{Channel State Information - Reference Signal}
\newacronym{dc}{DC}{Dual Connectivity}
\newacronym{dce}{DCE}{Direct Code Execution}
\newacronym{dci}{DCI}{Downlink Control Information}
\newacronym{dl}{DL}{Downlink}
\newacronym{dmr}{DMR}{Deadline Miss Ratio}
\newacronym{dmrs}{DMRS}{DeModulation Reference Signal}
\newacronym{e2e}{E2E}{End-to-End}
\newacronym{ecn}{ECN}{Explicit Congestion Notification}
\newacronym{edf}{EDF}{Earliest Deadline First}
\newacronym{enb}{eNB}{evolved Node Base}
\newacronym{epc}{EPC}{Evolved Packet Core}
\newacronym{es}{ES}{Edge Server}
\newacronym{fdma}{FDMA}{Frequency Division Multiple Access}
\newacronym{fdd}{FDD}{Frequency Division Duplexing}
\newacronym[firstplural=Radio Access Technologies (RATs)]{rat}{RAT}{Radio Access Technology}
\newacronym{fs}{FS}{Fast Switching}
\newacronym{ftp}{FTP}{File Transfer Protocol}
\newacronym{gnb}{gNB}{Next Generation Node Base}
\newacronym{harq}{HARQ}{Hybrid Automatic Repeat reQuest}
\newacronym{hetnet}{HetNet}{Heterogeneous Network}
\newacronym{hh}{HH}{Hard Handover}
\newacronym{hol}{HOL}{Head-of-Line}
\newacronym{ia}{IA}{Initial Access}
\newacronym{imt}{IMT}{International Mobile Telecommunication}
\newacronym{iot}{IoT}{Internet of Things}
\newacronym{los}{LOS}{Line of Sight}
\newacronym{lte}{LTE}{Long Term Evolution}
\newacronym{m2m}{M2M}{Machine to Machine}
\newacronym{mac}{MAC}{Medium Access Control}
\newacronym{mc}{MC}{Multi-Connectivity}
\newacronym{mcs}{MCS}{Modulation and Coding Scheme}
\newacronym{mec}{MEC}{Mobile Edge Cloud}
\newacronym{mi}{MI}{Mutual Information}
\newacronym{mimo}{MIMO}{Multiple Input, Multiple Output}
\newacronym{mmwave}{mmWave}{millimeter wave}
\newacronym{mr}{MR}{Maximum Rate}
\newacronym{mss}{MSS}{Maximum Segment Size}
\newacronym{mtd}{MTD}{Machine-Type Device}
\newacronym{mtu}{MTU}{Maximum Transmission Unit}
\newacronym{nfv}{NFV}{Network Function Virtualization}
\newacronym{nlos}{NLOS}{Non Line of Sight}
\newacronym{nr}{NR}{New Radio}
\newacronym{ofdm}{OFDM}{Orthogonal Frequency Division Multiplexing}
\newacronym{pdcch}{PDCCH}{Physical Downlonk Control Channel}
\newacronym{pdcp}{PDCP}{Packet Data Convergence Protocol}
\newacronym{pdsch}{PDSCH}{Physical Downlink Shared Channel}
\newacronym{pdu}{PDU}{Packet Data Unit}
\newacronym{pf}{PF}{Proportional Fair}
\newacronym{pgw}{PGW}{Packet Gateway}
\newacronym{phy}{PHY}{Physical}
\newacronym{pbch}{PBCH}{Physical Broadcast Channel}
\newacronym[plural=\gls{mme}s,firstplural=Mobility Management Entities (MMEs)]{mme}{MME}{Mobility Management Entity}
\newacronym{prb}{PRB}{Physical Resource Block}
\newacronym{pss}{PSS}{Primary Synchronization Signal}
\newacronym{pucch}{PUCCH}{Physical Uplink Control Channel}
\newacronym{pusch}{PUSCH}{Physical Uplink Shared Channel}
\newacronym{rach}{RACH}{Random Access Channel}
\newacronym{ran}{RAN}{Radio Access Network}
\newacronym{red}{RED}{Robotics Emergency Deployment}
\newacronym{rf}{RF}{Radio Frequency}
\newacronym{rlc}{RLC}{Radio Link Control}
\newacronym{rlf}{RLF}{Radio Link Failure}
\newacronym{rrc}{RRC}{Radio Resource Control}
\newacronym{rrm}{RRM}{Radio Resource Management}
\newacronym{rr}{RR}{Round Robin}
\newacronym{rs}{RS}{Remote Server}
\newacronym{rsrp}{RSRP}{Reference Signal Received Power}
\newacronym{rss}{RSS}{Received Signal Strength}
\newacronym{rtt}{RTT}{Round Trip Time}
\newacronym{rw}{RW}{Receive Window}
\newacronym{rx}{RX}{Receiver}
\newacronym{sa}{SA}{standalone}
\newacronym{sack}{SACK}{Selective Acknowledgment}
\newacronym{sap}{SAP}{Service Access Point}
\newacronym{sch}{SCH}{Secondary Cell Handover}
\newacronym{scoot}{SCOOT}{Split Cycle Offset Optimization Technique}
\newacronym{sdma}{SDMA}{Spatial Division Multiple Access}
\newacronym{sinr}{SINR}{Signal to Interference plus Noise Ratio}
\newacronym{sm}{SM}{Saturation Mode}
\newacronym{snr}{SNR}{Signal to Noise Ratio}
\newacronym{son}{SON}{Self-Organizing Network}
\newacronym{ss}{SS}{Synchronization Signal}
\newacronym{srs}{SRS}{Sounding Reference Signal}
\newacronym{sss}{SSS}{Secondary Synchronization Signal}
\newacronym{tb}{TB}{Transport Block}
\newacronym{tcp}{TCP}{Transmission Control Protocol}
\newacronym{tdd}{TDD}{Time Division Duplexing}
\newacronym{tdma}{TDMA}{Time Division Multiple Access}
\newacronym{tfl}{TfL}{Transport for London}
\newacronym{tm}{TM}{Transparent Mode}
\newacronym{trp}{TRP}{Transmitter Receiver Pair}
\newacronym{tti}{TTI}{Transmission Time Interval}
\newacronym{ttt}{TTT}{Time-to-Trigger}
\newacronym{tx}{TX}{Transmitter}
\newacronym{ue}{UE}{User Equipment}
\newacronym{ul}{UL}{Uplink}
\newacronym{uml}{UML}{Unified Modeling Language}
\newacronym{um}{UM}{Unacknowledged Mode}
\newacronym{utc}{UTC}{Urban Traffic Control}
\newacronym{vm}{VM}{Virtual Machine}
\newacronym{rsrq}{RSRQ}{Reference Signal Received Quality}
\newacronym{rssi}{RSSI}{Received Signal Strength Indicator}
\newacronym{crs}{CRS}{Cell Reference Signal}
\newacronym{comp}{CoMP}{Coordinated Multi-Point}
\newacronym{cran}{C-RAN}{Cloud \acrlong{ran}}
\newacronym{ca}{CA}{Carrier Aggregation}
\newacronym{cco}{CC}{Carrier Component}
\newacronym{nsa}{NSA}{Non Stand Alone}
\newacronym{embb}{eMBB}{Enhanced Mobility Broadband}
\newacronym{bsr}{BSR}{Buffer Status Report}
\newacronym{srb}{SRB}{Service Radio Bearer}
\newacronym{scm}{SCM}{Spatial Channel Model}
\newacronym{sctp}{SCTP}{Stream Control Transmission Protocol}
\newacronym{mptcp}{MPTCP}{Multi-path TCP}
\newacronym{ietf}{IETF}{Internet Engineering Task Force}
\newacronym{os}{OS}{Operating System}
\newacronym{tls}{TLS}{Transport Layer Security}
\newacronym{rfc}{RFC}{Request for Comments}
\newacronym{http}{HTTP}{HyperText Transfer Protocol}
\newacronym{nat}{NAT}{Network Address Translation}
\newacronym{api}{API}{Application Programming Interface}
\newacronym{rto}{RTO}{Retransmission Timeout}
\newacronym{psc}{PSC}{Public Safety Communication}
\newacronym{rpgm}{RPGM}{Reference Point Group Mobility}
\newacronym{ic}{IC}{Incident Command}
\newacronym{rsu}{RSU}{Road Side Unit}
\newacronym{uav}{UAV}{unmanned aerial vehicle}
\newacronym{usv}{USV}{Unmanned Surface Vehicle}
\newacronym{uas}{UAS}{Unmanned Aerial System}
\newacronym{iab}{IAB}{Integrated Access and Backhaul}
\newacronym{qoe}{QoE}{Quality of Experience}
\newacronym{ssim}{SSIM}{Structural Similarity Index}
\newacronym{psnr}{PSNR}{Peak Signal to Noise Ratio}
\newacronym{bs}{BS}{Base Station}
\newacronym{mu}{MU}{Multiple User}
\newacronym{ag}{AG}{Air-to-Ground}
\newacronym{af}{AF}{Array Factor}
\newacronym{ula}{ULA}{Uniform Linear Array}
\newacronym{upa}{UPA}{Uniform Planar Array}
\newacronym{lcs}{LCS}{Local Coordinate System}
\newacronym{psd}{PSD}{Power Spectral Density}
\newacronym{vq}{VQ}{vector quantization}
\newacronym{a2g}{A2G}{air-to-ground}
\newacronym{em}{EM}{electromagnetic}
\newacronym{vae}{VAE}{variational autoencoder}

\def\bb0{{\mathbb{0}}}

\def\bb{{\boldsymbol{b}}}

\def\b0{{\boldsymbol{0}}}

\def\b{{\mathrm{b}}}

\def\r0{{\mathbf{0}}}

\def\bsf0{{\bm{\mathsf{0}}}}

\def\N0{{N_{\mathrm{0}}}}

\def\bsf{{\boldsymbol{s}_\mathrm{f}}}

\newcommand{\be}{\begin{equation}}
\newcommand{\ee}{\end{equation}}
\newcommand{\bal}{\begin{align}}
\newcommand{\eal}{\end{align}}

\def\dB {{|_{\mathrm{\scriptscriptstyle dB}}}}

\def\ISDs   {{\mathsf{ISD}_{\sf s}}}
\def\ISDd   {{\mathsf{ISD}_{\sf d}}}

\makeatletter

 \let\oldforeign@language\foreign@language
 \DeclareRobustCommand{\foreign@language}[1]{%
   \lowercase{\oldforeign@language{#1}}}

\usepackage[font=small]{caption}

\usepackage{dblfloatfix}

\def\nb0{{\mathbf{0}}}
\def\nb1{{\mathbf{1}}}

\def\N{\sigma^2}

\makeatother

\IEEEoverridecommandlockouts

\begin{document}
%
\title{Millimeter-Wave UAV Coverage \\ in Urban Environments}
%

\author{
%
\IEEEauthorblockN{Seongjoon Kang$^{\dagger}$ \quad Marco Mezzavilla$^{\dagger}$ \quad Angel Lozano$^{\flat}$  \quad Giovanni Geraci$^{\flat}$ \\ \quad William Xia$^{\dagger}$ \quad Sundeep Rangan$^{\dagger}$   \quad Vasilii Semkin$^{\sharp}$ \quad Giuseppe Loianno$^{\dagger}$} 
\IEEEauthorblockA{$^{\dagger}$NYU Tandon School of Engineering, Brooklyn, NY, USA}
\IEEEauthorblockA{$^{\sharp}$VTT Technical Research Centre of Finland Ltd, Finland}
\IEEEauthorblockA{$^{\flat}$Univ. Pompeu Fabra, Barcelona, Spain}
\thanks{S. Rangan, W. Xia, S. Kang, and M. Mezzavilla were supported by NSF grants  1302336,  1564142,  1547332, and 1824434,  NIST, SRC, and the industrial affiliates of NYU WIRELESS. A.~Lozano and G. Geraci were supported by ERC grant 694974, by MINECO's Project RTI2018-101040, by the Junior Leader Fellowship Program from ``la Caixa" Banking Foundation, and by the ICREA Academia program. The work of V. Semkin was supported by the Academy of Finland.}
}

\maketitle

\begin{abstract}
With growing interest in mmWave connectivity for unmanned aerial vehicles (UAVs), a basic question is whether networks intended for terrestrial service can provide sufficient aerial coverage as well. To assess this possibility in the context of urban environments,
extensive system-level simulations are conducted using a generative channel model recently proposed by the authors.
It is found that standard downtilted base stations at street level, deployed with typical microcellular densities, can indeed provide satisfactory UAV coverage. Interestingly, this coverage is made possible by a conjunction of antenna sidelobes and strong reflections. As the deployments become sparser, the coverage is only guaranteed at progressively higher UAV altitudes. The incorporation of base stations dedicated to UAV communication, rooftop-mounted and uptilted, would strengthen the coverage provided their density is comparable to that of the standard deployment, and would be instrumental for sparse deployments of the latter.


\end{abstract}


\IEEEpeerreviewmaketitle

\section{Introduction}
\label{sec:intro}

Communication with unmanned aerial vehicles (UAVs) is a promising enhancement to 5G networks \cite{3GPP36777,GerGarAza2021,ZenGuvZha2020,SaaBenMoz2020,GerGarLin2019,FotQiaDin2019, MozSaaBen2018, ZenLyuZha2019, GerGarGal2018,GarGerLop2019,LinWirEul2019,AzaRosPol2019, NguAmoWig2018}.
Many compelling use cases for UAV communication demand consistently high bit rates: over 100~Mbps for things like 8K video broadcasting, critical missions, or surveillance; over 1~Gbps for wireless backhaul enabling radio access via UAV hotspots, say for temporary crowded events or disaster relief \cite{3GPP22125}. 


Millimeter wave (mmWave) frequencies provide an opportunity to meet the demands of these high-data-rate UAV applications
\cite{6824752,6736746}.  These frequencies offer massive bandwidths and, due
to directionality, are more immune to interference than their sub-6-GHz counterparts \cite{GerGarGal2018}.
In addition, aerial links, particularly at high altitudes, have pronounced likelihood
of line-of-sight (LOS) coverage and can avoid the blockage issues experienced 
by terrestrial mmWave networks
\cite{zhang2019research}.
Also, relative to a typical 
smartphone, UAVs have significantly higher power budgets and form factors to accommodate
the antenna arrays and the signal processing needed by mmWave transceivers \cite{xia2019uav,SemKanHaa2021}.


As current and envisioned 5G mmWave deployments primarily target ground users, the question arises as to whether such infrastructure suffices to provide satisfactory aerial coverage; if that is not the case, mobile operators will have to deploy dedicated infrastructure to enable aerial connectivity. This question, while pressing, had not been addressed because of the lack of a mmWave aerial channel model.

Capitalizing on a generative channel model recently developed \cite{XiaRanMez2020a} and freely available in open source \cite{mmw-github},
the present paper investigates this aerial coverage problem.
Atop this powerful channel model, capable of representing highly complex multipath propagation scenarios, 
a system-level simulator has been built; this simulator is also freely available \cite{sj-github}.
For the 3GPP antenna patterns \cite{xia2019uav} and under 5G urban deployment assumptions
\cite{3GPP36777}, 
it is shown that:
\begin{itemize}
\item At typical microcellular densities, 5G mmWave networks can indeed
provide satisfactory coverage to aerial links. This result is surprising because
mmWave antennas are highly directional and standard base stations---gNBs in 5G parlance---are downtilted.  Importantly, such coverage turns out to be possible via a combination of antenna sidelobes and strong reflections.
\item As their density declines, the coverage from standard gNBs becomes progressively less robust. For low-density deployments, dedicated gNBs mounted on rooftops and uptilted could substantially enhance coverage.
\end{itemize}

\section{Generative Channel Model}

\begin{figure}[!t]
\centering
\includegraphics[width=0.99\columnwidth]{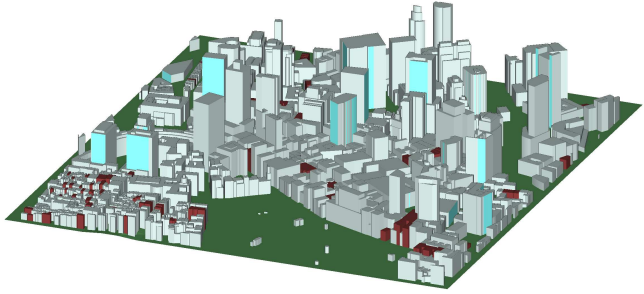}
\caption{3D map of a section of Boston employed in the model training.}
\label{fig:remcom}
\end{figure}

As will be seen, aerial coverage depends in a complex manner 
on the angular and power distributions of signal paths along with the antenna patterns at the UAV and gNB. To capture these distributions accurately, this work employs
a recently developed data-driven generative channel model 
\cite{XiaRanMez2020,XiaRanMez2020a,mmw-github}.  
This model 
operates in a two-stage fashion:
\begin{itemize}
\item The first stage accepts as inputs the location of the UAV relative to the gNB, and the type of gNB in question.
Given the UAV-gNB separation and the type of gNB,
the model determines the probability of the link being in one of three states: LOS (meaning an LOS path does exist, possibly in addition to other paths),
NLOS (meaning no LOS path exists, only other paths), or outage (meaning no paths are available).  
\item The second stage in the channel model generates, conditioned on the link state (LOS/NLOS/outage), 
a random number of paths along with the angles, delays, and strengths of the corresponding multipath components. This stage is powered by a variational autoencoder, which is a form of generative model suitable for neural networks.
\end{itemize}
Both stages can be trained to match the conditional distribution of path parameters
found in any environment of interest. For this work, the model was trained on over 20000 links for a section of Boston (see Fig. \ref{fig:remcom}) simulated by the Wireless InSite ray tracing tool \cite{Remcom}.

\section{Simulation Framework}
\label{sec:sim}

\begin{table}[!b]
  \begin{center}
    \caption{Simulation parameters.}
    \label{tab:table1}
    \begin{tabular}{|l|l|} 
    \hline
      \textbf{Parameter} & \textbf{Value}\\
      \hline
      Area (\SI{}{\meter}$^2$)  &  $1000 \times 1000 $ \\
      Minimum distance between gNBs (\SI{}{\meter}) &  $10$ \\
      Minimum distance between UAVs and gNBs (\SI{}{\meter}) &  $10$ \\
      $\ISDs$ (\SI{}{\meter}) & $400$, $200$\\
      $\ISDd$ (\SI{}{\meter}) & $800, 400, 200 $\\
      Height of standard gNBs (\SI{}{\meter}) &  Unif[2,5] \\
      Height of dedicated gNBs (\SI{}{\meter}) & Unif[10,30] \\
      Bandwidth (\SI{}{\MHz}) & $400 $ \\
      Frequency (\SI{}{\GHz}) & $28 $ \\
      gNB noise figure (\SI{}{\dB}) & $6$\\
      UAV TX power (\SI{}{dBm}) & $23$\\
      gNB antenna configuration & $8\times8$ URA\\
      UAV antenna configuration & $4\times4$ URA\\
      3GPP Vertical half-power beamwidth ($\theta_{\text{3dB}}$)  & $65^\circ$\\
      3GPP Horizontal half-power beamwidth ($\phi_{\text{3dB}}$) & $65^\circ$ \\
      \hline
    \end{tabular}
  \end{center}
\end{table}

\begin{figure}[!t]
\centering
\includegraphics[width=0.99\columnwidth]{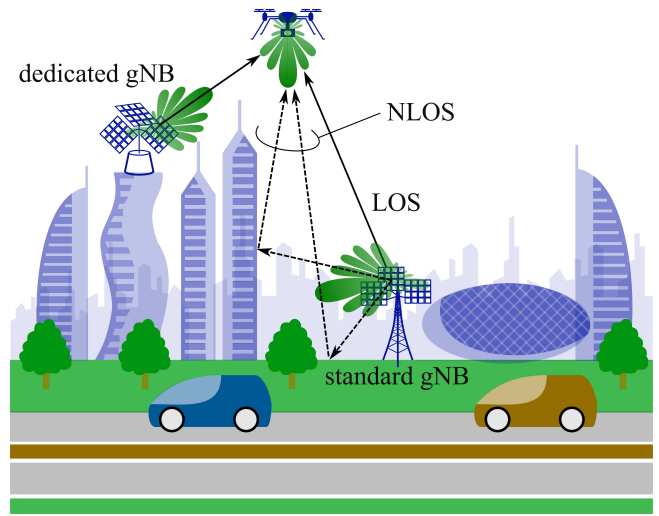}
\caption{Urban deployment featuring standard and dedicated gNBs.}
\label{fig:network_topology}
\end{figure}

Our study is conducted on a wrapped-around \SI{1}{\km} $\times$ \SI{1}{\km} universe
featuring a network of \emph{standard} street-level gNBs deployed uniformly at random subject to a certain average intersite distance ($\ISDs$) and, to prevent singularities, subject also to a minimum intersite distance; each gNB spawns three sectors, with a certain downtilt. On top of this, a second network of \emph{dedicated} gNBs can be optionally deployed, only on rooftops and with an uptilt. 
The transmitting UAVs are also distributed uniformly, on planes at specific heights.
The overall setting is illustrated in Fig.~\ref{fig:network_topology} and detailed in Table \ref{tab:table1}. The Python-based system-level simulator developed to obtain the coverage insights reported in this paper is freely available  to facilitate reproducibility and research follow-ups \cite{sj-github}.

\begin{figure}[!t]
\centering
\includegraphics[width=0.99\columnwidth]{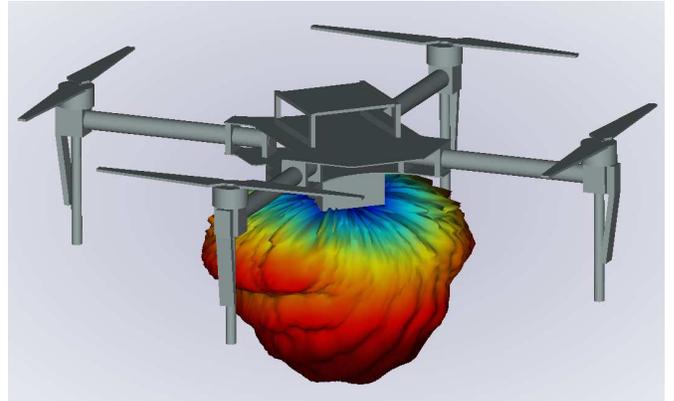}
\caption{Radiation pattern of a simulated patch antenna mounted on the bottom of a drone.}
\label{fig:uav_patch}
\end{figure}

\begin{figure}[!t]
\centering
\includegraphics[width=0.99\columnwidth]{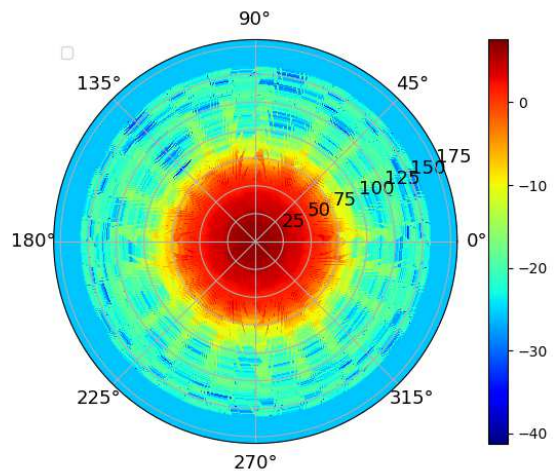}
\caption{Simulated UAV antenna field pattern.}
\label{fig:antenna_pattern}
\end{figure}

\begin{figure*}[tpbh]
\centering
  \subfloat[SNR distribution for UAVs at 30 m.]{\includegraphics[width = 0.99\columnwidth, trim = 1cm 0 1cm 0, clip]{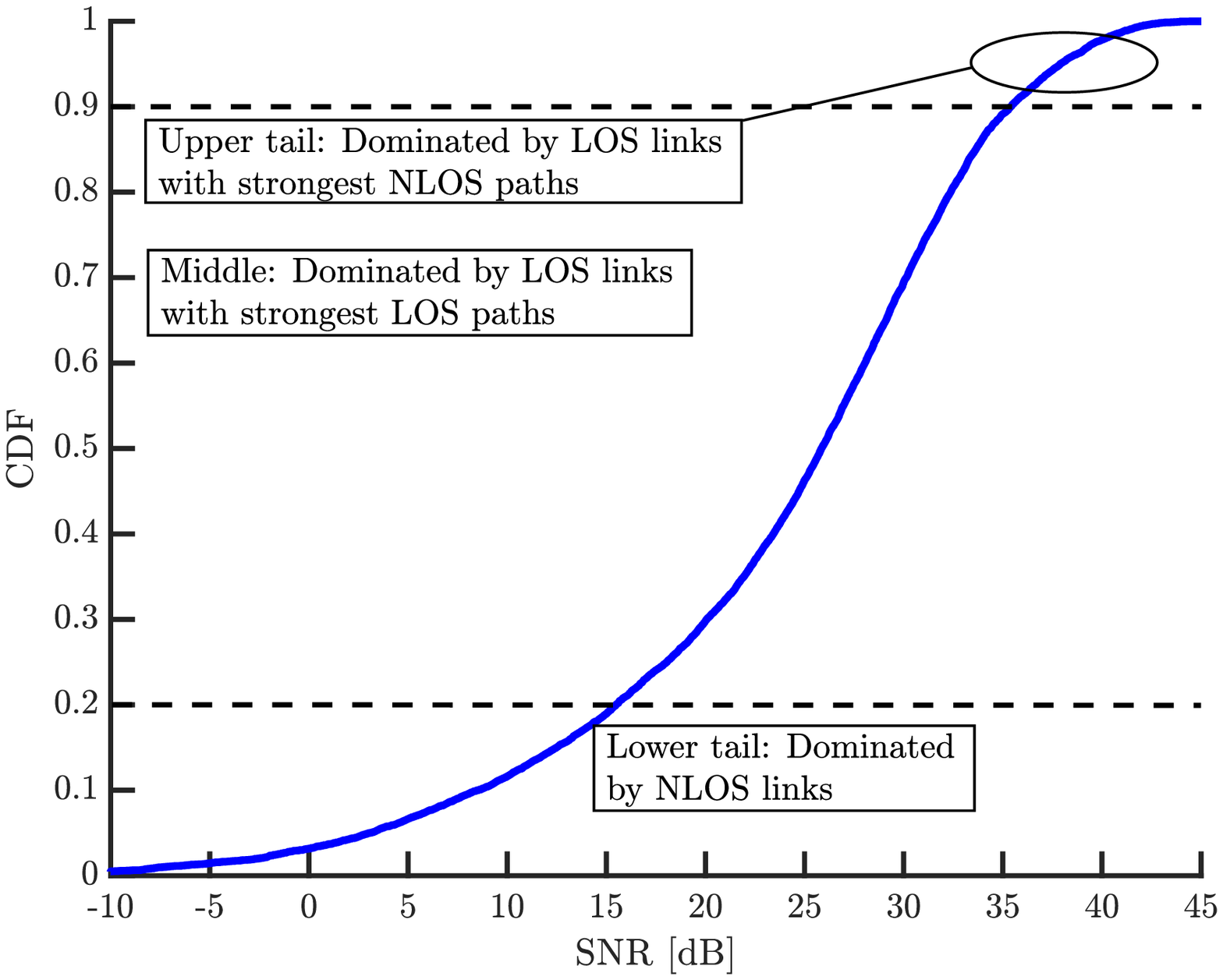}}
  \subfloat[Link states for UAVs at 30 m.]{\includegraphics[width = 0.95\columnwidth, trim = 1cm 0 1cm 0, clip]{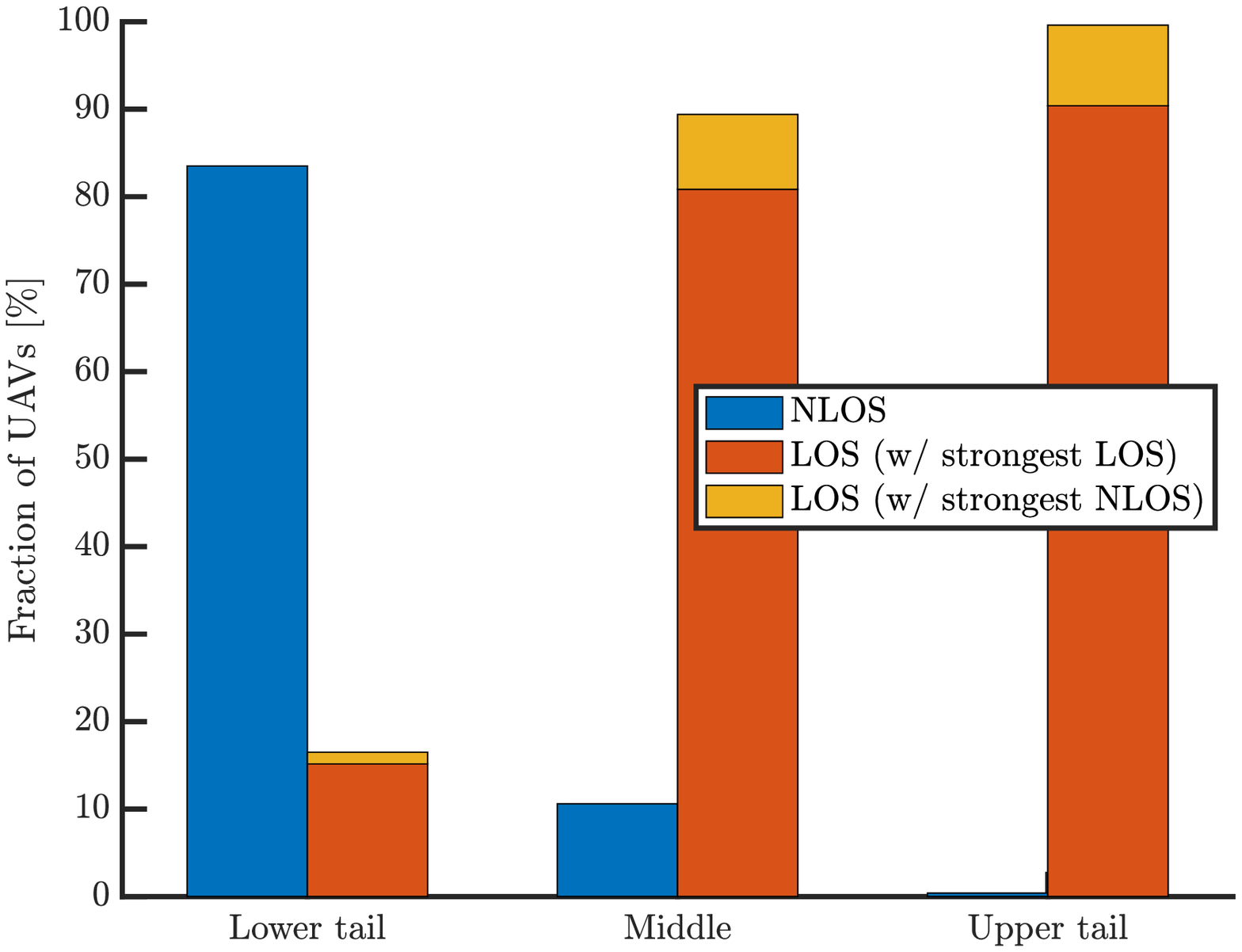}}
  \\
  \subfloat[SNR distribution for UAVs at 120 m.]{\includegraphics[width = 0.99\columnwidth, trim = 1cm 0 1cm 0, clip]{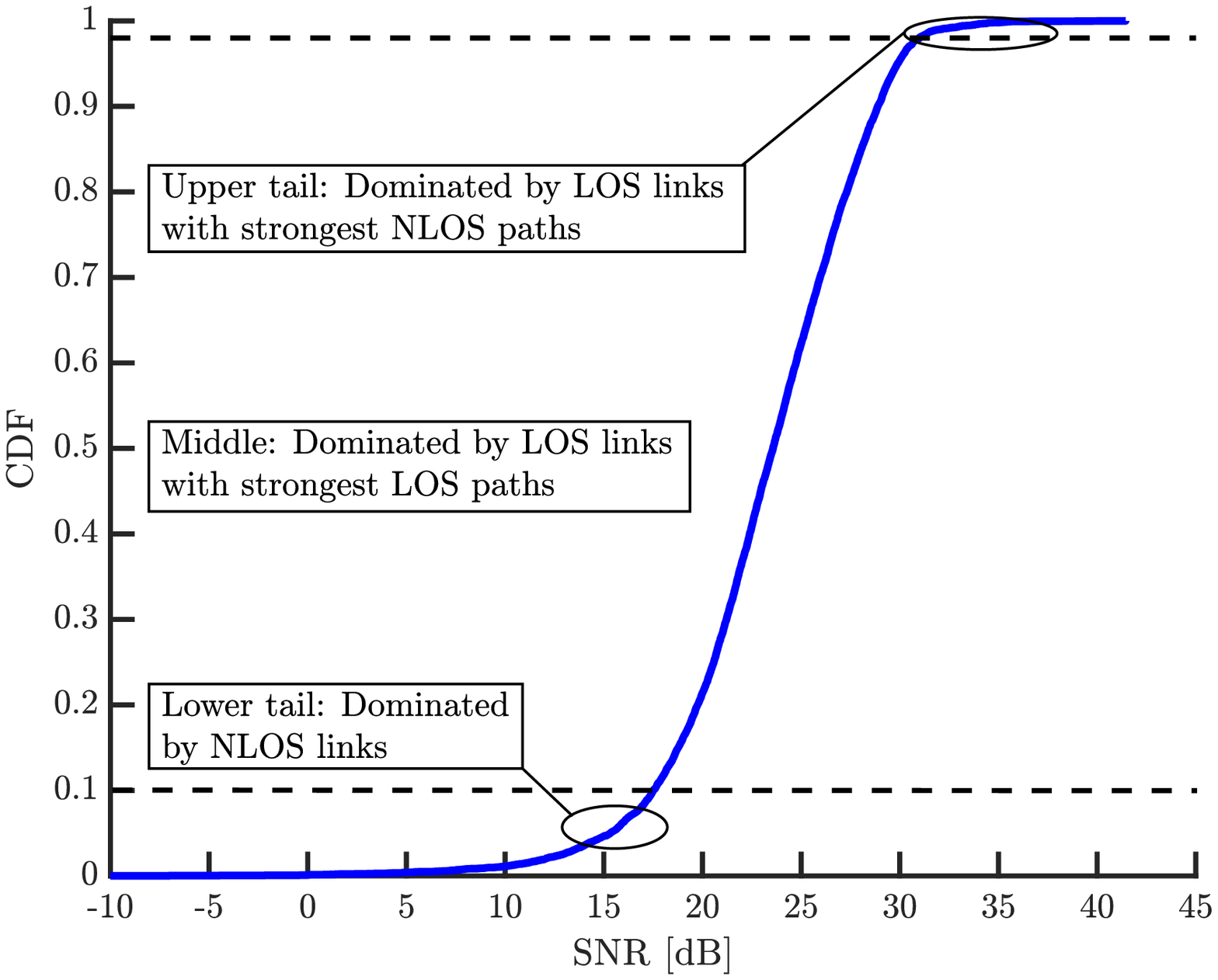}}
  \subfloat[Link states for UAVs at 120 m.]{\includegraphics[width = 0.99\columnwidth, trim = 1cm 0 1cm 0, clip]{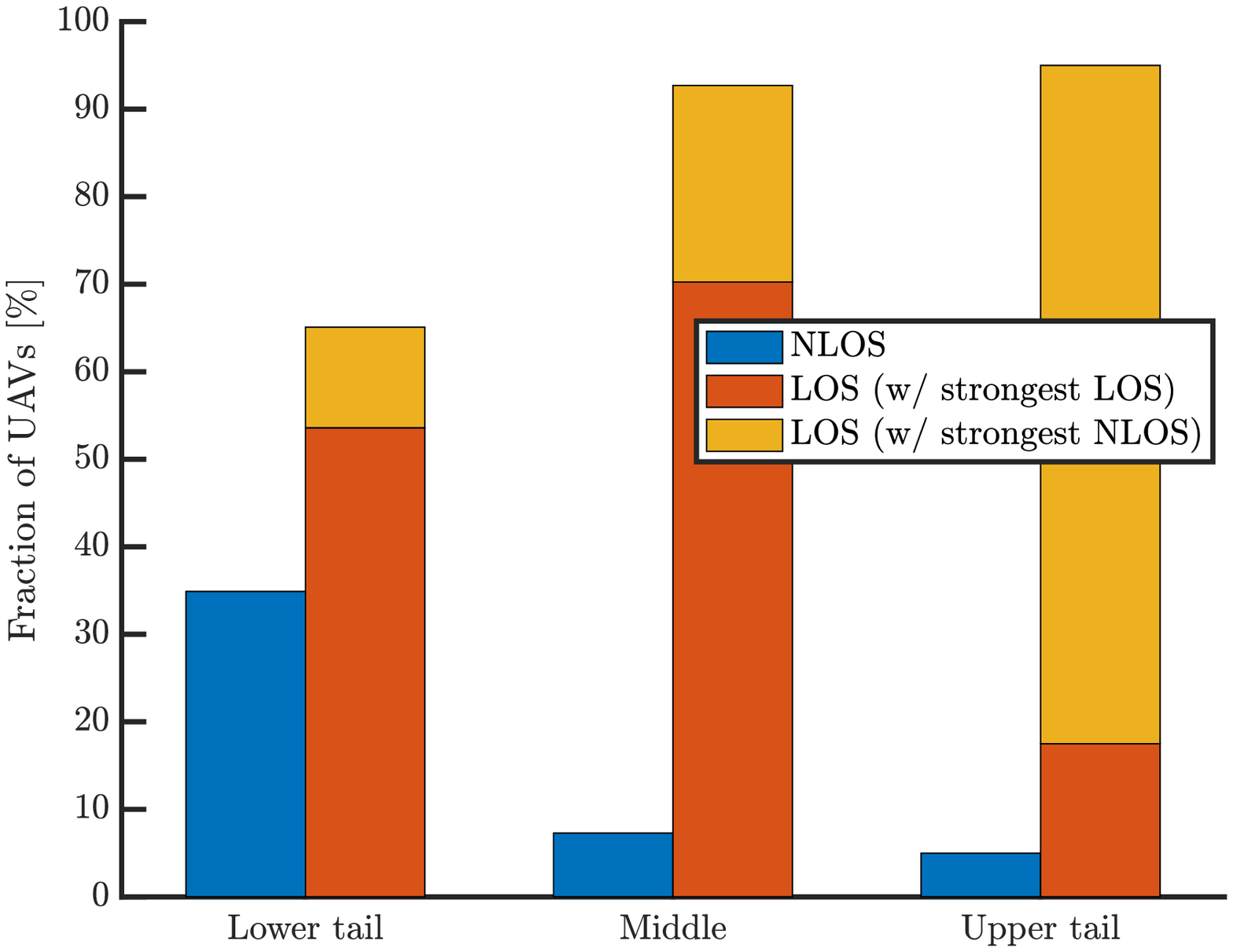}}
   \caption{SNR and link state distributions for $\ISDs=200$ m with UAV altitudes of 30 and 120 m.}
   \label{fig:snr_regimes}
\end{figure*}

Each gNB features an $8 \times 8$ uniform rectangular array (URA) per sector while each UAV is equipped with a single $4 \times 4$ URA.
The antennas of standard and dedicated gNBs are, respectively, downtilted by $-12^{\circ}$ and uptilted by $45^{\circ}$. In turn, the UAV antennas are pointed downwards ($-90^{\circ}$). 

From the locations of the gNBs and UAVs, and the type of gNBs, the generative model
generates the random multipath channels for each gNB-UAV pair.  
We focus on the uplink (UAV $\rightarrow$ gNB) since this is the power-limited link.
From the multipath channel and antenna array assumptions, the uplink 
SNR on each gNB-UAV link can be straightforwardly calculated \cite{heath2018foundations}. 
Specifically, each randomly generated
channel consists of $L$ paths.  For each path $\ell$,
the model outputs the angles of arrival (AOA) in elevation and azimuth,
$\theta_{\ell}^{\sf rx}$ and $\phi_{\ell}^{\sf rx}$, the angles of departure (AOD), 
$\theta_{\ell}^{\sf tx}$ and $\phi_{\ell}^{\sf tx}$, and the power gain, $G_\ell$. We adopt the gNB antenna pattern specified by 3GPP \cite{3GPP36873} and the simulated UAV antenna pattern shown in Fig.~\ref{fig:uav_patch} with polar representation given in Fig.~\ref{fig:antenna_pattern} \cite{SemKanHaa2021}. 
This pattern is from a patch antenna element simulated in a high frequency structure simulator using geometric optics at \SI{28}{GHz}.
The element is mounted facing vertically downwards on a 3D model of a drone frame composed of carbon fiber material.
The patterns on the transmit and receive arrays, along with these AOAs and AODs, yield the transmit and receive element gains for
path $\ell$, denoted by $A^{\sf tx}_\ell$ and $A^{\sf rx}_\ell$.
Standard long-term beamforming is applied, whereby the transmit and receive beamforming vectors align with the maximal-eigenvalue eigenvector of the channel covariance \cite{akdeniz2014millimeter,heath2018foundations}.
This 
altogether yields the SNR for each link, and then each UAV 
connects to the gNB offering the highest SNR.


\section{Coverage with Standard $\mathrm{gNBs}$ }

Let us begin by considering a deployment reliant exclusively on standard gNBs.
The goal in this section is to assess the extent to which such a deployment can provide satisfactory aerial coverage.

With the SNR taken as a proxy for coverage, Fig.~\ref{fig:snr_regimes} presents the distribution of the SNR with $\ISDs=$ \SI{200}{m} with UAV altitudes of 30 and \SI{120}{m}. At least $95\%$ of UAVs at \SI{30}{m} have an SNR$>$ \SI{0}{dB}, testifying to an acceptable coverage at low altitudes.  Moreover, the coverage improves with altitude.  By \SI{120}{m}, almost all UAVs achieve in excess of \SI{15}{dB}.

\begin{figure}[t]
\centering
    \includegraphics[width = 0.9\columnwidth, trim = 1cm 0 1cm 0, clip]{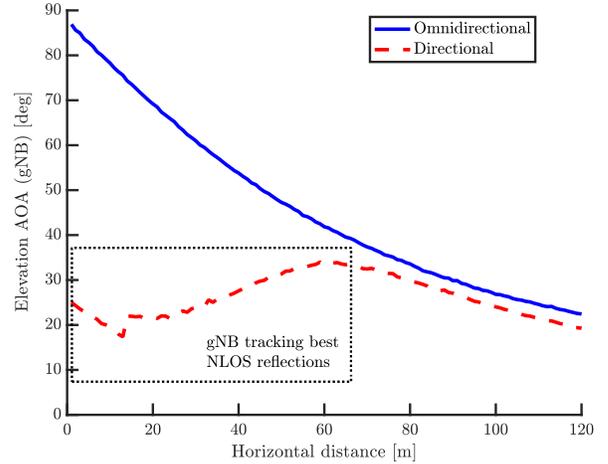}
    \caption{Elevation AOA for the strongest path at the gNB with both omnidirectional and directional ($\theta_{\text{3dB}} = 65\degree$, $\phi_{\text{3dB}} = 65\degree$, maximum gain 8 dBi) antenna elements. The UAV travels horizontally at a fixed altitude of 60 meters. For each horizontal displacement, the results are averaged over 10000 realizations.}
  \label{fig:reflection_gains}
\end{figure}

The result is surprising given the downtilted directional nature of the antennas
at the gNBs, with a front-to-back ratio of \SI{30}{dB}.
To understand how UAVs still enjoy satisfactory coverage, Fig.~\ref{fig:snr_regimes}
identifies three distinct SNR regions and then examines the distribution
of link states in each such region.  We observe that:
\begin{itemize}
\item The lower tail corresponds predominantly to NLOS links.
\item The middle section is dominated by LOS links whose LOS component is in turn dominant.
\item The upper tail is mostly constituted by LOS links, but with an NLOS link being the strongest
link after accounting for the antenna gains.
\end{itemize}
Thus, in the lower tail, which is what determines the coverage, the connectivity rests mostly on NLOS communication, meaning through a conjunction of gNB antenna sidelobes and reflected paths. Both these ingredients turn out to play a substantial role. To illustrate this point, Fig.~\ref{fig:reflection_gains} depicts the elevation AOA for the strongest path at the gNB
as the UAV travels from the vertical of that gNB to a horizontal distance of \SI{500}{m} at an altitude of \SI{60}{m}. The result is averaged over 10000 channel realizations, and presented for antenna elements having the 3GPP antenna pattern as well as for omnidirectional elements. The comparison reveals that the gNB leverages NLOS paths that happen to be stronger than the LOS, especially when the UAV is horizontally close. Conversely, without antenna directivity, the gNB would merely track the weaker LOS path. 

\begin{figure}[tpbh]
\centering
    \includegraphics[width=0.9\columnwidth, trim = 0.3cm 0 0.8cm 0, clip]{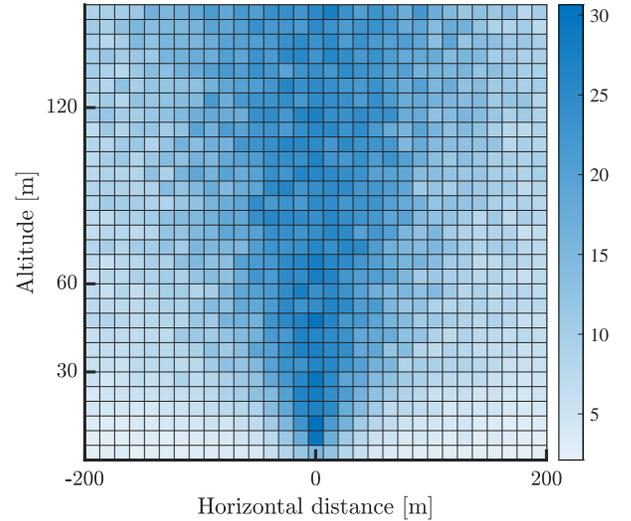}
    \caption{Standard deviation of the elevation AOA (in degrees) at the gNB, as a function of the horizontal distance and altitude of the UAV relative to that gNB, which is positioned at (0,0).}
    \label{fig:std_aoa}
\end{figure}

As the UAV altitude increases, the LOS probability grows, but NLOS paths continue to play an important SNR-enhancing role. This phenomenon is evidenced by Fig.~\ref{fig:std_aoa}, which visually shows how the range of elevation AOAs expands with the altitude, reinforcing the probability that an NLOS path impinges from a direction with substantial antenna gain.

Altogether, a standard 5G mmWave deployment suffices for UAV coverage when $\ISDs=\SI{200}{m}$.
Then, as $\ISDs$ extends beyond this value, coverage decays for low-altitude UAVs, and a progressively higher minimum altitude is required for service. For $\ISDs=\SI{400}{m}$, for instance, roughly $10\%$ of UAVs at \SI{30}{m} turn out to experience negative-dB SNRs, indicating that coverage is no longer guaranteed at this altitude.


\begin{figure}[tpbh]
    \centering
	\includegraphics[width=0.97\columnwidth, trim = 0.2cm 0 0.5cm 0, clip]{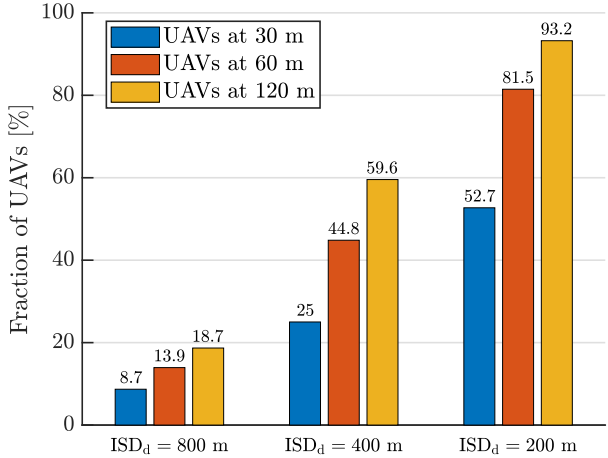}
	\caption{Fraction of UAVs that choose to connect to dedicated gNBs when $\ISDs=\SI{200}{m}$, parameterized by $\ISDd$ and by the UAV altitude.}
	\label{fig:fractionOfconnectedToBSs}
\end{figure}

\begin{figure}[tpbh]
\centering
\subfloat[UAV altitude: \SI{30}{m}.]{\includegraphics[width=0.97\columnwidth, trim = 0.5cm 0 0.0cm 0, clip]{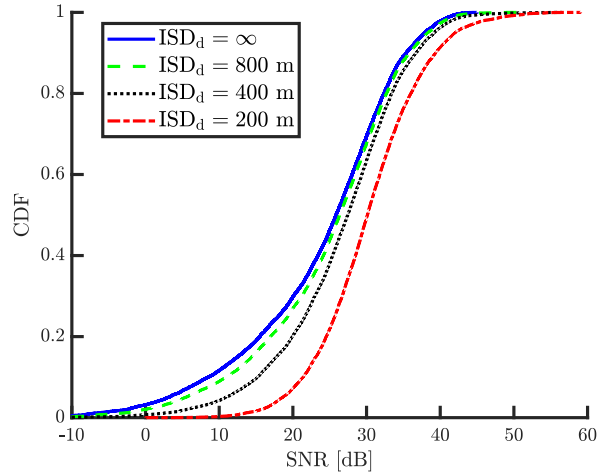}}\\
\subfloat[UAV altitude: \SI{60}{m}.]{\includegraphics[width=0.97\columnwidth, trim = 0.5cm 0 0.0cm 0, clip]{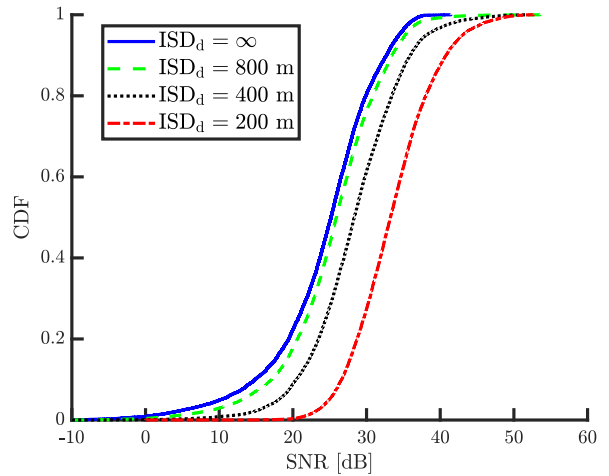}}\\
\subfloat[UAV altitude: \SI{120}{m}.]{\includegraphics[width=0.97\columnwidth, trim = 0.5cm 0 0.0cm 0, clip]{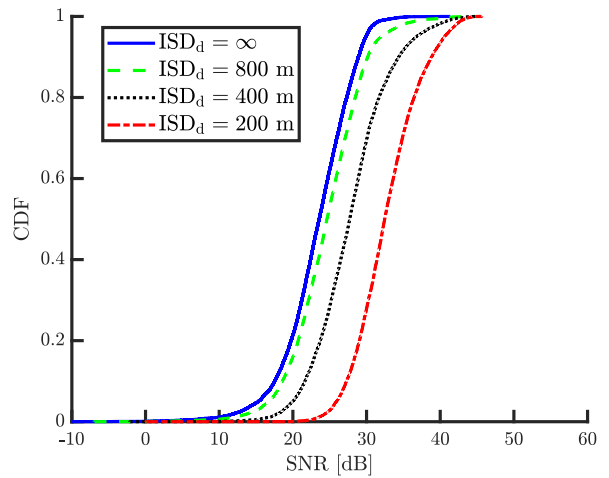}}
\caption{SNR distributions for UAVs at different heights randomly distributed in urban scenarios where connectivity is provided by a combination of standard gNBs ($\ISDs= $200) and dedicated gNBs ($\ISDd= $200, 400, 800).}
\label{fig:snrdistributions}
\end{figure}

\section{Coverage Enhancement with Dedicated $\mathrm{gNBs}$}

Having characterized the coverage for a standard deployment, let us now quantify the impact of incorporating dedicated gNBs with $\ISDd \geq \ISDs$. Fig. \ref{fig:fractionOfconnectedToBSs} shows the fraction of UAVs that choose to connect to such dedicated gNBs, for $\ISDd= $200, 400, or \SI{800}{m}, rather than connect to standard gNBs with $\ISDs= $\SI{200}{m}. 


This fraction is presented for various UAV altitudes, indicating that it is the UAVs at higher altitudes that exhibit increased preference for the dedicated gNBs. This is toned down at lower altitudes, but even then dedicated gNBs are overall favored when their density equals that of the standard ones. As one would expect, the relevance of dedicated gNBs dwindles as they become sparser.

The ensuing SNR distributions, with each UAV connecting to its preferred gNB, either standard or dedicated, are presented in Fig. \ref{fig:snrdistributions} for altitudes of 30, 60 and \SI{120}{m}. (The leftmost curves for 30 and \SI{120}{m} coincide with those in Fig.~\ref{fig:snr_regimes}, when the dedicated gNBs are absent.)
The SNR improvement with dedicated gNBs is pronounced, especially in the lower tail and for higher-altitude UAVs, provided those dedicated gNBS are as dense as their standard counterparts. With sparser dedicated gNBs, the improvement weakens, becoming anecdotal for $\ISDd = 4 \ISDs$.

\begin{figure}[h]
\centering
    \includegraphics[width = 0.97 \columnwidth, trim = 0.5cm 0 0.0cm 0, clip]{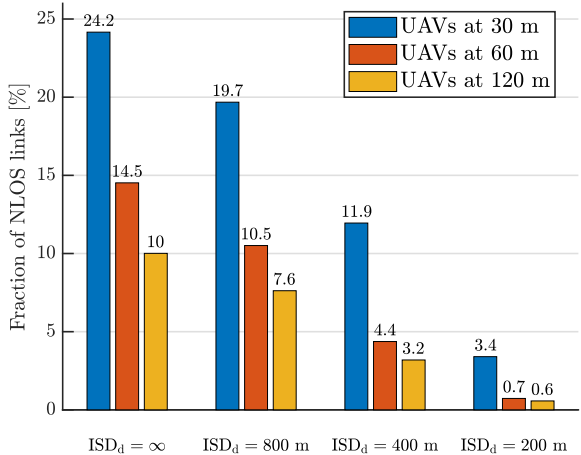}
    \caption{Fraction of NLOS links at each height for $\ISDs$ = $\SI{200}{m}$, as a function of $\ISDd$.}
    \label{fig:linkratio}
\end{figure}

Another perspective on the effects of deploying dedicated gNBs is provided in Fig. \ref{fig:linkratio}, which shows the fraction of links that are NLOS.
The addition of dedicated gNBs drastically curbs that fraction at all the considered UAV altitudes and, with $\ISDd=\ISDs$, a vast majority of UAVs enjoy LOS connectivity to their serving gNBs.

\section{Conclusion}
\label{sec:conclusion}

On the basis of a generative-model trained to recreate the propagation conditions in
the city of Boston, this paper has established that a standard 5G  mmWave deployment, intended for ground users, suffices to provide satisfactory aerial coverage; qualitatively similar observations have been drawn with the generative model trained on sections of Boston, Moscow, Tokyo, and Beijing.
The potential enhancements provided by additional dedicated gNB, uptilted and rooftop-mounted, have also been quantified.
More in detail, the takeaways of our study are as follows:
\begin{itemize}
    \item For UAVs at 120~m and higher, standard deployments provide strong coverage thanks to a combination of gNB antenna sidelobes and strong NLOS paths produced by reflections. Due to the downtilt preventing strong antenna gains in the LOS direction, the LOS path may be overpowered by some of those NLOS reflections.
    \item Standard infrastructure can also cover UAVs at 30--120~m very well, as long as the deployment is dense enough, namely for $\ISDs \leq 200$.
    \item Deploying additional dedicated gNBs enhances the SNR, but the effect is only pronounced if their density is comparable to that of the standard gNBs. If the standard deployment is sparser than $\ISDs=200$ m, dedicated gNBs are likely to be instrumental to ensure coverage.
\end{itemize}

We note that all these observations have been collected for long-term beamforming, and the ensuing conclusions would only be reinformed with more sophisticated beamforming procedures reliant on instantaneous channel-state information at the trasmitter.

Also noteworthy is that, while the reported results correspond to static simulations,
multipath dynamics can be reproduced from the outputs of the generative channel model, opening the door to characterizing the impact of channel estimation and imperfect beamforming.

Further planned extensions of this work include accounting for intercell interference, spatial multiplexing, and the interplay with ground users, with the ultimate goal of evaluating not only the coverage but further the \emph{capacity} that urban mmWave deployments can provide to UAVs, as well as designing UAV offloading, mobility management, and power control strategies for heterogeneous standard-plus-dedicated deployments.



\bibliographystyle{IEEEtran}

\end{document}